\documentclass{article}

\usepackage{PRIMEarxiv}

\usepackage[utf8]{inputenc} % allow utf-8 input
\usepackage[T1]{fontenc}    % use 8-bit T1 fonts
\usepackage{hyperref}       % hyperlinks
\usepackage{url}            % simple URL typesetting
\usepackage{booktabs}       % professional-quality tables
\usepackage{amsfonts}       % blackboard math symbols
\usepackage{nicefrac}       % compact symbols for 1/2, etc.
\usepackage{microtype}      % microtypography
\usepackage{lipsum}
\usepackage{fancyhdr}       % header
\usepackage{graphicx}       % graphics
\usepackage{multirow}
\usepackage{threeparttable}
\graphicspath{{media/}}     % organize your images and other figures under media/ folder

\title{Learning-by-teaching with ChatGPT: The effect of teachable ChatGPT agent on programming education
}
\pdfoutput=1 

\author{
  Angxuan Chen \\
  Department of Educational Technology\\ Peking University\\ Beijing, China\\
  \texttt{angxuan.chen@stu.pku.edu.cn} \\
   \And
  Yuang Wei\thanks{Corresponding author: philrain@foxmail.com} \\
  Lab of Artificial Intelligence for Education\\ East China Normal University\\ Shanghai, China\\
  \texttt{philrain@foxmail.com} \\
  \And
  Huixiao Le \\
  Department of Educational Technology\\ Peking University\\ Beijing, China\\
  \texttt{interesting@pku.edu.cn} \\
  \And
  Yan Zhang \\
  Department of Computer Science and Technology\\ Tsinghua University\\ Beijing, China\\
  \texttt{yan-zhan23@mails.tsinghua.edu.cn} \\
}

\begin{document}
\maketitle

\begin{abstract}
This study investigates the potential of using ChatGPT as a teachable agent to support students' learning by teaching process, specifically in programming education. While learning by teaching is an effective pedagogical strategy for promoting active learning, traditional teachable agents have limitations, particularly in facilitating natural language dialogue. Our research explored whether ChatGPT, with its ability to engage learners in natural conversations, can support this process. The findings reveal that interacting with ChatGPT improves students' knowledge gains and programming abilities, particularly in writing readable and logically sound code. However, it had limited impact on developing learners' error-correction skills, likely because ChatGPT tends to generate correct code, reducing opportunities for students to practice debugging. Additionally, students' self-regulated learning (SRL) abilities improved, suggesting that teaching ChatGPT fosters learners' higher self-efficacy and better implementation of SRL strategies. This study discussed the role of natural dialogue in fostering socialized learning by teaching, and explored ChatGPT's specific contributions in supporting students' SRL through the learning by teaching process. Overall, the study highlights ChatGPT's potential as a teachable agent, offering insights for future research on ChatGPT-supported education.
\end{abstract}

% keywords can be removed

\section{Introduction}
Learning by Teaching (LBT) is an innovative instructional method that has garnered significant attention in educational research in recent years. LBT requires learners not only to express and reorganize existing knowledge but also to actively engage in reflective knowledge construction. This approach to knowledge construction goes beyond simple knowledge transmission, encouraging learners to create and refine knowledge that is useful for learning by forming deeper explanations, analogies, and connections across broader contexts\cite{chase2009teachable,chi2014icap}. Most studies investigating learning-by-teaching, entailed interacting with an audience \cite{duran2017learning,roscoe2007understanding}. Considering the limited availability of peers for teaching exercises, past research has looked into utilizing AI as ``teachable agents" to support in the LBT method. These agents are programmed to pose questions and make demands, encouraging learners to teach them. When used as an instructional activity, students are often explicitly told to adopt a teaching role by communicating distinct information on the subject matter to a fictitious fellow student with low levels of prior knowledge \cite{Lachner2019Timing}. Through this interactive teaching process, learners not only instruct the agents but also deepen their own understanding and mastery of the subject matter \cite{silvervarg2021teachable}. Empirical results supported the effectiveness of teachable agents on achieving deeper learning and transfer \cite{kim2016based}, emotional regulation \cite{han2021teachable} and motivation \cite{kim2013gendered}. 

However, the development and implementation of teachable agents are not without challenges. One of the primary limitations is the technical capability of these agents. Most teachable agents developed thus far can only handle predefined problems and are limited in their interaction capabilities, often restricted to non-verbal responses or simplistic interactions such as button clicking or filling in blanks (e.g, \cite{biswas2016design, pareto2009learning}). This limitation stems from current gaps in technology that limit these agents from processing and responding to natural language inputs in a way that mimics human-like interactions. As a result, these agents are often unable to initiate or sustain interactions that are perceived as socially rich and engaging by human standards. As previous studies pointed out \cite{lachner2022learning}, learning by teaching is inherently social. The limited interaction capabilities of these agents hinder learners' engagement and reflection. For instance, research has shown that the effectiveness of learning by teaching varies depending on the user’s belief about whether they are interacting with a human or a computer agent. Even when the agents' responses are identical, students tend to offer less detailed self-explanations and respond with brief keywords when interacting with an agent, compared to when they believe they are interacting with a human \cite{Ogan2012Oh}.

Currently, the rapid development of generative AI, particularly models like ChatGPT, has introduced a natural dialogue method that are capable of engaging in much deeper and more sophisticated conversations with learners\cite{liu2023summary,lo2023impact} . Unlike traditional teachable agents that are limited by predefined problem sets and simplistic interaction modes, ChatGPT and similar generative AI models can process natural language inputs and generate human-like responses \cite{fui2023generative}. Research has shown that learners often perceive ChatGPT as a human-like agent, especially when it assists with learning tasks \cite{luo2024How, markel2023gpteach}. However, there is limited research on the effects of using ChatGPT as a teachable agent and the teaching strategies employed in such environments. This study investigated the impact of ChatGPT-based teachable agents in a programming learning task, and examined learners' knowledge gains, programming skills, and self-regulated learning (SRL) abilities. The goal was to explore ChatGPT's potential in enhancing the learning-by-teaching process.

\section{Background and Related Works}\label{sec2}

\subsection{Learning by Teaching}

As a widely recognized instructional method, LBT not only encourages learners to express and reorganize their existing knowledge but also actively engages them in reflective knowledge construction. Unlike simple knowledge transmission, LBT promotes the creation and refinement of understanding by guiding learners to develop deeper explanations, analogies, and connections across diverse contexts \cite{chase2009teachable,chi2014icap,duran2017learning,roscoe2007understanding}.
At the core of LBT is the idea that teaching prompts learners to move beyond merely expressing what they already know. Through reflection and reorganization, they develop a deeper understanding and construct a new knowledge system. This knowledge system in LBT is crucial, it not only draws on learners' existing knowledge but also challenges them to extend this knowledge beyond the provided materials, building more complex and nuanced cognitive structures\cite{nasir2014knowledge}. 
On the other hand, teaching itself is a highly reflective process. Learners must continually monitor and adjust their understanding to ensure that they can clearly convey knowledge to others. This reflective process naturally enhances learners' metacognitive abilities, making them more conscious and proactive in monitoring their cognitive processes and learning progress\cite{spruce2015teacher,kinnebrew2013investigating}. 
In the teaching process, learners become aware of their own knowledge gaps through teaching others. These ``knowledge gaps" prompt them to take action and further refine their understanding, thereby strengthening their self-regulation\cite{biswas2004developing}. 
Lastly, LBT fosters a sense of responsibility and motivation in students, encouraging them to take greater ownership of planning and managing their learning tasks, which further promotes the development of both cognitive and metacognitive processes in self-regulation \cite{biswas2004incorporating,biswas2009promoting}.
The processes of self-reflection, knowledge restructuring, self-monitoring, and self-regulation that occur during learning fall under the scope of Self-Regulated Learning (SRL) \cite{puustinen2001models,zimmerman2002becoming}, and combining the SRL model with the LBT approach has been proven to be a more effective theoretical way to enhance learning outcomes \cite{biswas2005learning}.

Despite the theoretical potential of LBT, its practical application in classrooms faces notable challenges. First, LBT demands students to alternate between the roles of teacher and learner, which is time-consuming and may hinder its widespread adoption in classroom settings\cite{king1998mutual}. Second, studies suggest that while the tutor's learning is enhanced, the learner might not benefit as much, potentially creating an imbalance in learning outcomes\cite{king1998mutual,walker2014adaptive}. In response to these challenges, researchers have explored technological solutions, with Teachable Agents (TAs) emerging as a prominent approach to overcoming LBT’s limitations. TAs are virtual agents designed to learn from students' explanations and demonstrations, effectively acting as learners. By doing so, they provide scalability to LBT, as they are available at any time and help reduce psychological barriers like the fear of making mistakes or the pressure of immediate responses during teaching\cite{chase2009teachable,debbane2023learning}. TAs are constructed around key components such as knowledge representation, learning mechanisms, interactive interfaces, feedback systems, metacognitive abilities, and social interaction. These elements enable TAs to engage students in reflective learning processes, enhancing their academic performance, self-efficacy, and metacognitive skills\cite{chin2010preparing}. 

\subsection{Teachable Agents in Learning by Teaching}

More specifically, the positive impact of TAs on learning gains can be attributed to the fact that the core component of LBT is the peer learner \cite{debbane2023learning,matsuda2010tuning}, and technology-supported TAs eliminate the need for human peers, which are typically required in traditional LBT approaches \cite{matsuda2012studying}. TAs learn declarative and procedural knowledge from students' explanations and demonstrations, effectively acting as peer learners in LBT\cite{blair2007pedagogical}. For example, Jun et al. used Linear Kid as a peer in tutoring sessions and found that while varying levels of prior knowledge influenced learning gains, the problem-solving process itself did not differ significantly across conditions\cite{jun2003facilitating}. Similarly, Mioduser et al. used the DynaLearn platform and showed that short-term interventions through graphical manipulation had minimal impact on learning outcomes\cite{bredeweg2009dynalearn}, whereas more interactive methods, such as games\cite{pareto2014teachable} or 3D virtual environments\cite{zhao2012learning}, significantly enhanced conceptual knowledge acquisition. 
Although these studies have achieved good results in their respective goals, they still share some common issues. For instance, the capabilities of TAs are limited, as they primarily provide support at the cognitive level and struggle to enhance metacognitive skills\cite{bredeweg2009dynalearn}. Additionally, the interaction methods have significant limitations; even when using games and virtual environments, the high implementation cost remains a major concern\cite{pareto2014teachable,zhao2012learning}.

To address these issues, researchers have also conducted in-depth studies on strategies and effectiveness of TAs in enhancing metacognitive skills. In an experiment with Betty's Brain\cite{biswas2005learning}, Biswas (2010) explored the impact of TAs on metacognitive development. The study compared three interventions—Learning by Teaching (LBT), Self-Regulated Learning (SRL), and an Intelligent Coaching System (ICS)—with controlled meta-tutor feedback. Results showed that while SRL and LBT students performed equally well on tests, SRL students created more accurate concept maps, and both outperformed ICS in test scores and map accuracy. Notably, SRL students, who received metacognitive strategy feedback, exhibited more focused monitoring behaviors than LBT students, highlighting the value of cognitive and metacognitive feedback in improving learning outcomes\cite{biswas2010measuring,roscoe2013shallow}. Similarly, in Matsuda's studies with SimStudent\cite{matsuda2010simstudent}, the importance of self-explanation was examined. SimStudent solicits self-explanations by asking students to justify their feedback during tutoring sessions. Across two school studies, Matsuda found a strong correlation between the depth of self-explanation and student learning gains\cite{matsuda2013cognitive}. Additionally, in the aforementioned studies, whether using Betty's Brain or SimStudent, it is essential to consider the possible decision paths at each step and predefine fixed templates. Executing a specific task typically requires about 1,200 to 2,300 lines of Java code, and developing different topics demands a significant amount of time and effort \cite{matsuda2022teachable}.

Therefore, tools like Betty's Brain and SimStudent indeed offer a promising research direction with their interactive TA approaches, as well as methods to enhance metacognition through techniques like feedback on questions\cite{matsuda2013cognitive}. However, their high development costs still limit their broader application. Additionally, the pre-designed templates and decision paths may fall short in accommodating unexpected real-life scenarios, leading to interactions and feedback that resemble mechanical answer selection, lacking natural communication, which is not conducive to the development of metacognitive skills \cite{terrace2005metacognition}.

Reflecting on past studies, it's clear that current research on TAs has some limitations and gaps. Earlier work on metacognition didn't go deep enough and missed exploring how TAs affect more complex aspects\cite{blair2007pedagogical}. Also, past research aimed at improving metacognition was too strict and lacked a natural, personalized approach\cite{biswas2010measuring,roscoe2013shallow}. The structured interactions and analysis didn't do much to truly boost metacognition. Plus, creating these types of interactions takes a lot of time, is hard to adapt to new contexts, and doesn't work well in different situations. Overcoming these challenges is still a difficult task.

\subsection{Exploring the Integration of LLMs in the Learning by Teaching Approach}

Fortunately, the emergence of large language models (LLMs) seems to make it possible to develop smarter, more personalized TAs that can engage in more natural interactions. LLMs excel in contextual conversations\cite{ouyang2022training,ross2023programmer}, role-playing\cite{kong2023better,markel2023gpteach}, and learning from demonstrations\cite{brown2020language,radford2019language}, enabling more natural and credible tutoring interactions, such as writing and explaining code on demand. Unlike traditional LBT systems that rely on limited, predefined interactions\cite{biswas2001extending,leelawong2008designing,matsuda2010learning,pareto2011teachable}, LLM-powered agents allow flexible interactions, letting learners ask open-ended questions and explore various teaching methods, which enhances knowledge construction and metacognition\cite{aflalo2021students,chin2002student}. For instance, Markel et al.'s GPTeach project demonstrated how LLMs can simulate interactions between teaching interns and virtual students by setting roles and contexts within prompts\cite{markel2023gpteach}. This setup provided a more realistic training environment for interns, allowing them to interact with different types of virtual students. Such agents not only facilitate flexible teaching interactions but also allow learners to adapt their teaching strategies and improve based on feedback, making the process more dynamic and productive\cite{biswas2001extending,leelawong2008designing,matsuda2018metacognitive}.

The emergence of LLMs provides a powerful tool for creating new types of TAs. Their advanced language comprehension, flexible dialogue capabilities, and strong imitation skills enable them to simulate natural learning scenarios and adapt teaching strategies. These qualities make LLMs well-suited for the LBT approach. While research has shown the flexibility of LLMs in building TAs, studies on their actual impact on learning outcomes remain limited. Therefore, current research should focus on assessing the true effectiveness of LLM-based TAs and identifying areas where they can enhance learning. This calls for a deep exploration in a specific learning domain. Programming education, which involves the combined processes of knowledge acquisition and hands-on problem-solving, is an ideal area for such exploration. This domain allows for better diagnosis of various learner states, such as knowledge mastery, practical skills, and meta-cognitive attributes. Notably, in programming education, LLMs possess a broader and deeper understanding compared to typical learners. Their large-scale training across various programming languages, algorithms, and best practices enables them to act as human-like programming mentors\cite{kazemitabaar2024codeaid}. For instance, Kazemitabaar's study demonstrated how LLMs can assist students by providing instant feedback and step-by-step pseudocode explanations, offering enhanced support throughout the coding process\cite{kazemitabaar2023novices}. Unlike traditional rule-based systems, LLMs offer greater flexibility, allowing students to explore multiple strategies and teaching methods without being confined to rigid instructional paths\cite{pareto2011teachable,chen2023chatcot}. Moreover, LLM-powered TAs can have their cognitive behaviors, such as knowledge levels and questioning, precisely controlled\cite{packer2023memgpt,zhou2023recurrentgpt}, enhancing their reasoning abilities\cite{chen2023chatcot,kim2024language,lin2024swiftsage} and enabling a more tailored learning experience\cite{junprung2023exploring,park2023generative}. Therefore, considering the excellent capabilities of LLMs in programming education and the dual focus on measuring both knowledge and skills, we chose to delve deeper into the topic of programming education using the highly popular ChatGPT as our LLM tool.

This study will provide an in-depth analysis of ChatGPT-based TAs' impact on programming learning, focusing on the following research questions:

\begin{itemize}
    \item RQ1: What impact does the Teachable ChatGPT Agent have on learners' knowledge gains?
    \item RQ2: What impact does the Teachable ChatGPT Agent have on learners' programming skills?
    \item RQ3: What impact does the Teachable ChatGPT Agent have on learners' SRL abilities?
\end{itemize}

\section{Method}\label{sec3}

\subsection{Participants}
We recruited 41 university students from two universities (34 males and 7 females) with a background in computer science and who had self-reported they had fundamental coding ability in C++ programming. All participants were between the ages of 18 and 27, with an overall average age of 21.2 (SD = 2.4). The informed consent form, which was signed by all participants, stated that the data collected would be used solely for academic purposes and would not be disclosed to the public. Within the 41 participants, 20 were assigned to Experimental Group (EG), who had learning programming with a teachable ChatGPT agent, while the remaining 21 participants were assigned to Control Group (CG) who learning programming only with the instruction of online videos. 

\subsection{Learning materials}

\begin{figure}[tb]
    \centering
    \includegraphics[width=0.4\linewidth]{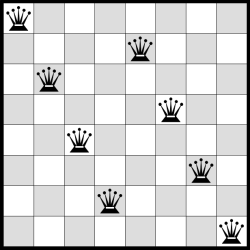}
    \caption{A sample solution of ``eight queens" puzzle that no two queens in the same row, column, or diagonal}
    \label{fig:sample_solution}
\end{figure}

The topic of our experimental learning material revolves around the ``eight queens" puzzle of AI algorithm learning, which includes algorithm knowledge learning and C++ programming learning. The ``eight queens" puzzle is a classic puzzle in computer science and mathematics. The challenge is to place eight queens on an 8x8 chessboard such that no two queens can attack each other. This means no two queens can be in the same row, column, or diagonal. A sample solution of ``eight queens” puzzle can be found in Figure.\ref{fig:sample_solution}. Here's a breakdown of the problem:

Chessboard Configuration: An 8x8 grid, like a standard chessboard.

Queen's Movement: A queen can move any number of squares vertically, horizontally, or diagonally.

Objective: Place eight queens on the board in such a way that no two queens can attack each other.

The participants need to calculate all possible placement coordinates of queens via programming. To assist participants in learning how to solve this problem through programming, we developed three online learning videos: (1) Fundamental knowledge learning Video: This video teaches participants the fundamental concepts of the backtracking algorithm used to solve this puzzle. (2) Eight Queen Puzzle Explanation Video: This video provides a detailed explanation of the ``Eight Queen" puzzle, outlining its rules and objectives. (3) Programming Solution Guide Video: In this video, a lecturer guided how to solve the puzzle with AI concept knowledge and programming, including a step-by-step code analysis with the backtracking algorithm. 

\begin{figure}[h!]
    \centering
    \includegraphics[width=0.7\linewidth]{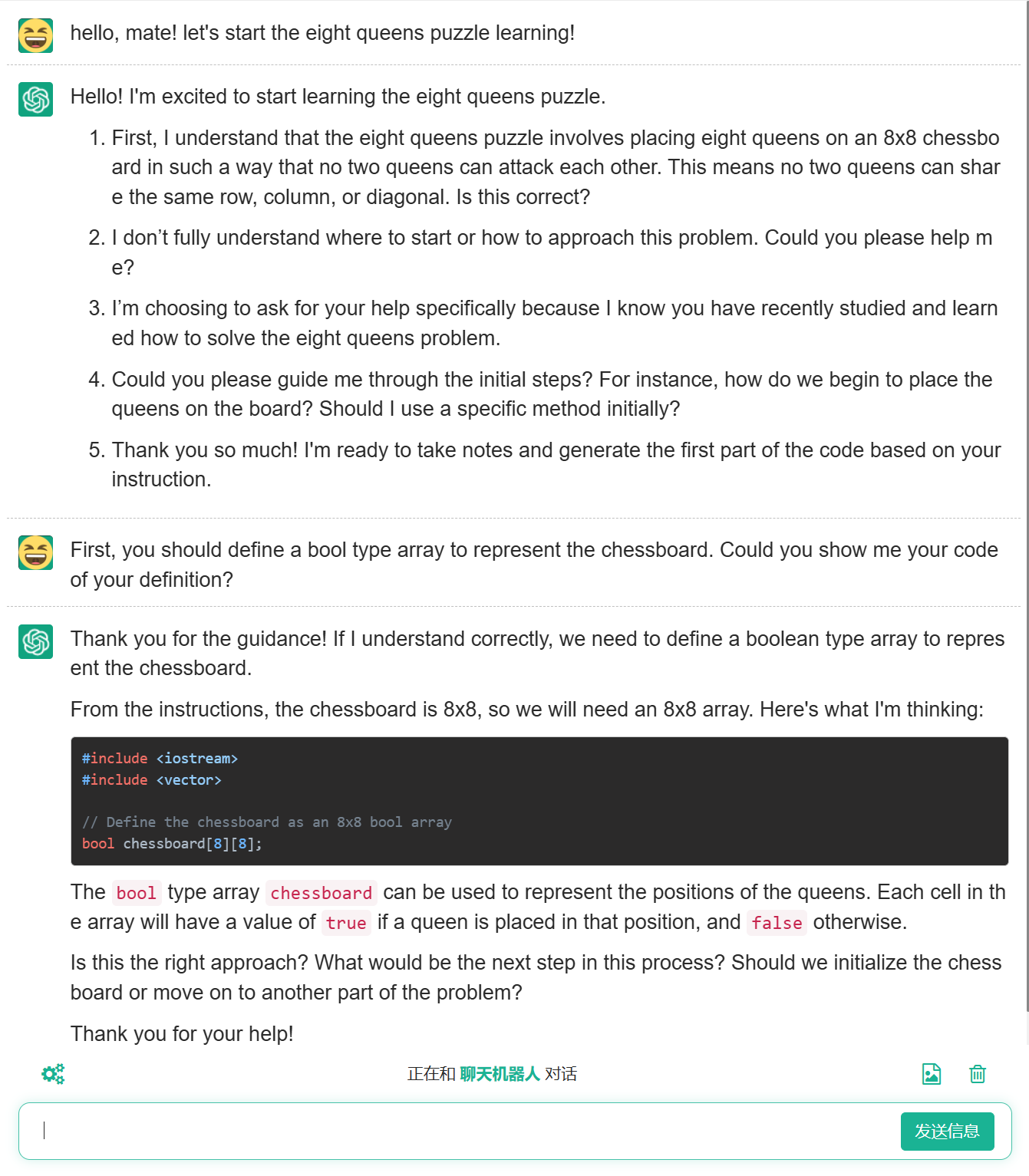}
    \caption{An example of learners' teaching in solving the ``eight queens" puzzle}
    \label{fig:example_teaching}
\end{figure}

After watching these three videos, participants in the experimental group (EG) learned to solve the ``Eight Queens" puzzle with a teachable agent we designed. We employed a prompt-based design approach that could modify the role of ChatGPT. To better design ChatGPT to act as a real student seeking help from teachers (participants), we adopted the theoretical help-seeking process model proposed by Gall (1981). This model describes five stages of the help-seeking process that students theoretically go through: (1) Awareness of the need for help. (2) Decision to seek help. (3) Identification of a potential help source. (4) Employment of strategies to elicit help. (5) Reactions to help-seeking attempts. Based on these stages, we structured ChatGPT’s responses to align with each stage by prompts, making ChatGPT a more realistic help-seeker. Specifically, we used the better version of ChatGPT, i.e., gpt-4, for our experiment, to ensure the prompt can be effectively understood by ChatGPT. The prompt used for this purpose can be found in the Appendix. In addition, As \cite{fui2023generative} mentioned, current ChatGPT may face the problem of hallucination, which means ChatGPT may generated some wrong content without control. However, when ChatGPT agent act as a teachable agent, those mistakes can be more like a beginner student act as and provide more opportunities that participants can teach with. Therefore, in our design, we did not chase the accuracy of ChatGPT agent in generating correct codes but used the original ChatGPT without other augmented techniques. An example of learners' teaching the ChatGPT teachable agent can be found in Figure.\ref{fig:example_teaching}.

Participants in EG were required to help the teachable ChatGPT to form correct codes that can solved the puzzle with previous instructional videos. Moreover, we developed an online judging platform that can automatically evaluate whether the participants' solutions of codes for the ``Eight Queens" puzzle are correct. The learning experiment was considered finished when participants successfully solved the puzzle with ChatGPT generated codes on the online platform or hit the time limit. Additionally, participants in the control group (CG) learned to solve the ``Eight Queens" puzzle only through those instructional videos, and they need to form the solution codes by themselves to pass the exam on the online judging platform.

\subsection{Procedure}
\begin{figure*}[tb]
    \centering
    \includegraphics[width=1\linewidth]{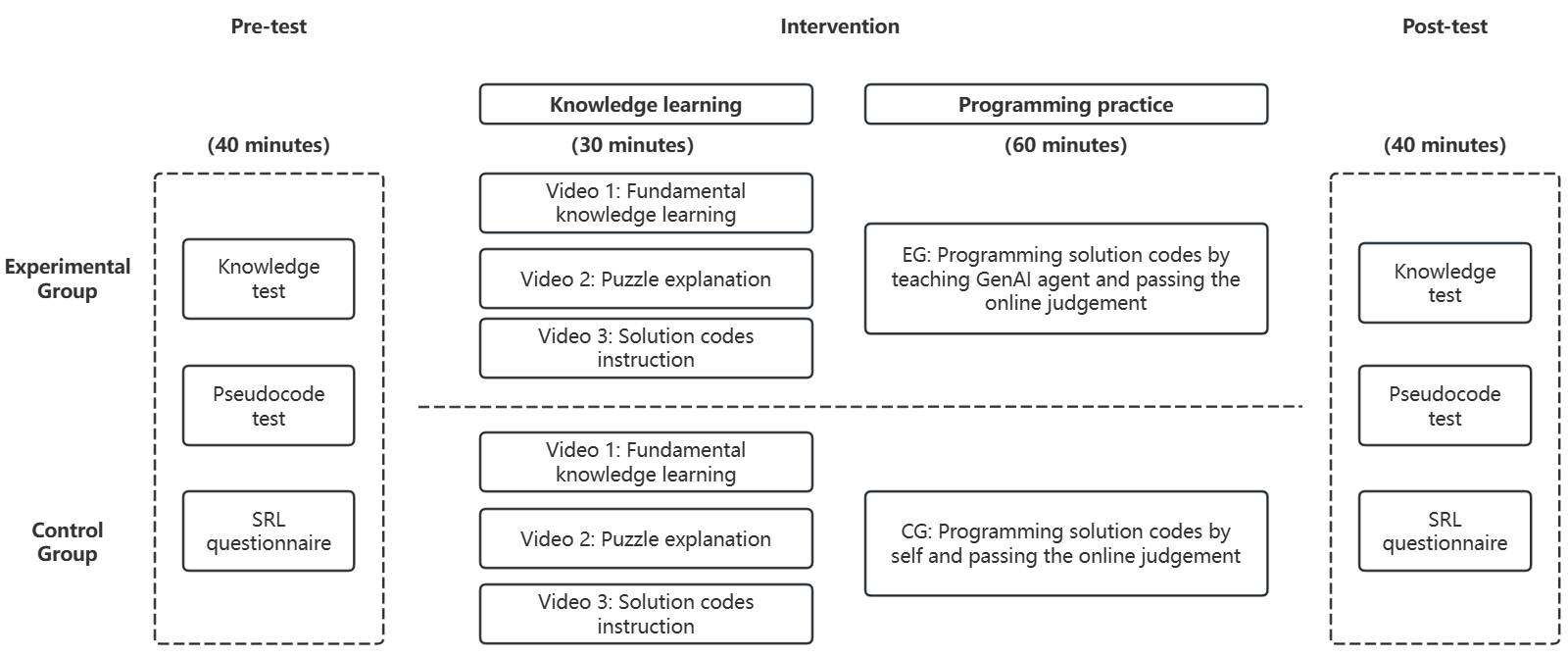}
    \caption{Procedure of the experiment}
    \label{fig:Procedure}
\end{figure*}

All the data for this study were collected in the online setting. It was clarified that the learning outcomes of this experiment would not impact their academic grades or performance. The complete experimental procedure is shown in Figure.\ref{fig:Procedure}. First, all the students were required to finished a pre-test online questionnaire including knowledge test and SRL test. Subsequently, the researchers introduced the experiment content to all participants, including introducing the three online videos and introducing the online judging platform for their codes’ submission. Second, all the participants were required to watch those three online videos to learn how to solve ``eight queen" puzzle in 30 minutes. Third, participants in EG were trying to form the solution codes with the ChatGPT teachable agent, while participants in CG were trying to form the solution codes by themselves in the local code editor. At this stage, participants were required to share their screens via online meeting software, with one of the authors supervising to ensure they followed the video instructions and did not engage in any cheating, such as searching for solution codes online. Participants had 1 hour to form correct codes by passing the online judging platform. Finally, all the participants were required to finish a post-test with knowledge test and SRL test.

\subsection{Measurements}

\subsubsection{Knowledge Test}

The knowledge test, designed by an experienced lecturer, consists of 15 single choice questions, focused on fundamental concepts of the backtracking algorithm and knowledge of ``eight queens" puzzle solution (e.g., What is the time complexity of the backtracking search algorithm for the ``eight queens" puzzle?). The test includes 5 easy, 5 medium, and 5 hard questions, with each question worth 1 point. The test was designed by an experienced lecturer and has been used for over 5 years. It’s been proven to be valid for measuring the knowledge learning of ``eight queens" puzzle.

\subsubsection{Pseudocode Test}

Pseudocode is an algorithm description language. The purpose of using pseudocode is to enable the described algorithm to be easily implemented in any programming language with a clear structure, simple code, good readability, and be similar to natural language. The well written pseudocode can represent participants’ programming level \cite{andrzejewska2016eye}. Our pseudocode test required participants to write the ``eight queens" puzzle solution with pseudocode. The pseudocodes were evaluated by two experienced scorers, focusing on three key aspects: clearness, correctness, and readability \cite{borstler2023investigating} (See detailed standard in Table \ref{tab1}). To ensure consistency in scoring, Spearman's correlation analysis was performed, revealing high-rate agreement across all three dimensions, with correlation coefficients of 0.833, 0.935, and 0.911, all statistically significant (p < 0.001). Following the scoring process, the scorers engaged in cross-checking and discussions to resolve any discrepancies, ultimately agreeing on the final evaluation scores.

\begin{table*}[!t]%
\centering %
\caption{Pseudocode test scoring standard\label{tab1}}%
\begin{tabular*}{\textwidth}{@{\extracolsep\fill}p{3cm}p{11cm}p{2cm}@{\extracolsep\fill}}
\toprule
\textbf{Dimension} & \textbf{Description}  & \textbf{Score}\\
\midrule
Clearness  & Clearness involves the logical flow and structure of the pseudocode, ensuring that the steps of the algorithm are straightforward and unambiguous. & 5\\
Correctness  & Correctness is determined by the ability of the pseudocode to solve the ``eight queens" puzzle as intended, ensuring that all conditions and rules of the puzzle are met without errors. & 5\\
Readability  & Refers to how easily the pseudocode can be read and interpreted, particularly in terms of formatting, consistency, and use of natural language-like constructs. Readability involves using clear variable names, consistent indentation, and comments or explanations that enhance understanding. & 5\\
\bottomrule
\end{tabular*}
\end{table*}

\subsubsection{SRL ability test}

The SRL ability was measured using an adapted version of the SRL questionnaire, which comprised 20 items rated on a five-point Likert scale. This questionnaire is derived from the Motivated Strategies for Learning Questionnaire (MSLQ) \cite{pintrich1990motivational}, a widely recognized SRL measurement grounded in social cognitive theory \cite{liao2024design}. The MSLQ includes two dimensions, namely motivation and strategy. The motivation section focuses on the values, expectations and influences of the learners and is typically divided into three sub-scales: self-efficacy, intrinsic value and test anxiety. The strategy section, on the other hand, emphasizes cognitive, metacognitive and resource management strategies and is commonly divided into the Cognitive Strategies Scale and the Self-Regulation Scale. Originally, the MSLQ consisted of 44 questions, but to reduce the burden on participants, it was condensed to 20 questions, and focusing on test anxiety, self-efficacy, and cognitive strategies. Some items were modified to align with the specific focus of this study. The final Cronbach's alpha for the measurement of test anxiety, self-efficacy, and cognitive strategies were 0.813, 0.791, and 0.912, which implied the high reliability of the measurement used in this study.

\section{Results}

\subsection{Knowledge test scores}

The difference in knowledge test scores between EG and CG was examined using an analysis of covariance (ANCOVA), with pre-test scores as the covariate and post-test scores as the dependent variable. Before the analysis, the basic assumptions of the data were examined first. The test for homogeneity of regression for the dependent variable in the EG and the CG depicted no significant difference with (F = 6.27, p > 0.05). Similarly, the Kolmogorov-Smirnov test revealed data distributions adhering to normality (p > .05). As a result, it was suitable to perform the ANCOVA test.

The ANCOVA results showed that the three groups were significantly different in knowledge test scores (F = 35.54, p < 0.05, $\eta$2 = 0.74) (as shown in Table \ref{knowledge_test_result}) that EG had a higher adjusted average score than CG (11.86 > 10.53). In other words, learners who learning AI knowledge with teaching the ChatGPT teachable agent have better knowledge gains than those who learning algorithm knowledge with online videos.

\begin{table*}[!t]%
\centering %
\caption{ANCOVA results of knowledge test scores\label{knowledge_test_result}}%
\begin{tabular*}{\textwidth}{@{\extracolsep\fill}llllllll@{\extracolsep\fill}}
\toprule
\textbf{Group} & \textbf{N}  & \textbf{Mean}  & \textbf{SD}  & \textbf{Adjusted Mean} & \textbf{Adjusted SE} & \textbf{F} & \textbf{$\eta$\textsuperscript{2} }\\
\midrule
EG & 20 & 11.90 & 2.09 &	11.86 & 0.32 & \multirow{2}*{35.54***} & \multirow{2}*{0.74}\\
CG & 21 & 10.76 &	1.02	& 10.53 & 0.24 \\
\bottomrule
\end{tabular*}

\end{table*}

\subsection{Programming test scores}
Before examining the impact of the ChatGPT teachable agent approach on participants' programming skills, Levene's test for equality of error variances showed comparable variance of the two groups (F = 2.93, p > 0.05), confirming that the variances were homogeneous. Additionally, the homogeneity of regression coefficients within the group was confirmed by the intra-group regression coefficient homogeneity test (F = 0.13, p > 0.05), allowing for the use of ANCOVA. The ANCOVA results (See Table \ref{programming_test_result}) revealed that participants in the experimental group, who learned programming with the teachable agent, scored significantly higher in code clearness and readability compared to the control group, which learned through online videos (F = 7.39, p < 0.01, $\eta^2$ = 0.37). However, there was no significant difference in code correctness between the two groups (F = 2.98, p > 0.05, $\eta^2$ = 0.19).

\begin{table*}[!t]%
\centering %
\caption{ANCOVA results of programming test scores\label{programming_test_result}}%
\begin{tabular*}{\textwidth}{@{\extracolsep\fill}lllllllll@{\extracolsep\fill}}
\toprule
\textbf{Dimension} & \textbf{Group} & \textbf{N}  & \textbf{Mean}  & \textbf{SD}  & \textbf{Adjusted Mean} & \textbf{Adjusted SE} & \textbf{F} & \textbf{$\eta$\textsuperscript{2} }\\
\midrule
\multirow{2}*{Clearness} & EG & 20 & 4.20	& 0.61	& 4.13 &	0.05 & \multirow{2}*{7.39***} & \multirow{2}*{0.37} \\
& CG & 21 & 3.29 & 0.90 & 3.17 & 0.01\\\multirow{2}*{Correctness} & EG & 20 & 3.81 & 0.96 & 3.79 & 0.10 & \multirow{2}*{2.98} & \multirow{2}*{0.19} \\
& CG & 21 & 3.90 & 1.47 & 3.82 & 0.03 \\\multirow{2}*{Clearness} & EG & 20 & 2.91 & 1.23 & 	2.95 & 0.03 & \multirow{2}*{4.32**} & \multirow{2}*{0.26} \\
& CG & 21 & 2.81 & 0.58 & 2.83 & 0.02 \\
\bottomrule
\end{tabular*}
\begin{tablenotes}
\item {\it Note.}: ***p<0.001; **p<0.01; *p<0.05
\end{tablenotes}
\end{table*}
\subsection{SRL abilities}

Before examining the impact of the ChatGPT teachable agent approach on participants’ SRL abilities, Levene's test for equality of error variances was conducted and revealed no significant difference in the pre-test scores between the two groups (F = 2.07, p > 0.05), indicating that the variances were homogeneous. Additionally, the test for the homogeneity of intra-group regression coefficients confirmed that the regression coefficients were consistent within each group (F = 2.753, p > 0.05), allowing for the use of ANCOVA. The ANCOVA results (See Table \ref{SRL_result}) demonstrated that the experimental group, which utilized the ChatGPT teachable agent, achieved significantly higher scores in self-efficacy (F = 37.26, p < 0.001, $\eta^2$ = 0.75), and cognitive strategies (F = 18.97, p < 0.001, $\eta^2$ = 0.61) compared to the control group. This suggests that learning with the ChatGPT teachable agent positively influences learners' self-efficacy and adoption of self-regulation strategies.

\begin{table*}[!t]%
\centering %
\caption{ANCOVA results of SRL abilities\label{SRL_result}}%
\begin{tabular*}{\textwidth}{@{\extracolsep\fill}lllllllll@{\extracolsep\fill}}
\toprule
\textbf{Dimension} & \textbf{Group} & \textbf{N}  & \textbf{Mean}  & \textbf{SD}  & \textbf{Adjusted Mean} & \textbf{Adjusted SE} & \textbf{F} & \textbf{$\eta$\textsuperscript{2} }\\
\midrule
 \multirow{2}*{Test anxiety}  & EG & 20 & 3.30 & 0.82 & 3.30 & 0.03 & \multirow{2}*{1.78} & \multirow{2}*{0.02} \\
 & CG &  21 & 3.46 & 0.74 & 3.44 & 0.09\\
  \multirow{2}*{Self-efficacy}  & EG & 20 & 3.88 & 0.59 & 3.81 & 0.13 & \multirow{2}*{37.26***} & \multirow{2}*{0.75} \\
 & CG &  21 & 3.54 & 0.82 & 3.50 & 0.14\\
   \multirow{2}*{Cognitive strategies}  & EG & 20 & 4.20 & 0.55 & 4.13 & 0.073 & \multirow{2}*{18.97***} & \multirow{2}*{0.61} \\
 & CG & 21 & 3.76 & 0.62 & 3.77 & 0.11\\
\bottomrule
\end{tabular*}
\begin{tablenotes}
\item {\it Note.}: ***p<0.001; **p<0.01; *p<0.05
\end{tablenotes}
\end{table*}

\section{Discussion}

\subsection{The effect of ChatGPT teachable agent on knowledge gains}

To address RQ1, we analyzed knowledge test scores and found that learners in the experimental group (EG), who taught a ChatGPT teachable agent, showed significantly greater knowledge gains compared to the control group (CG), who learned through online videos. This aligns with prior research, which indicates that learners who learning with the expectation of teaching later tend to achieve better outcomes than those who learning solely for a test (e.g., \cite{fiorella2013relative,fiorella2014role}). In our experiment, learners in the experimental group were informed to take responsibility for the solutions codes generated by the ChatGPT agent to pass the auto-judging platform. This accountability could motivate them to put in more effort, leading to enhanced learning outcomes. In the meantime, in the learning task, learners were required to guide the ChatGPT agent using only natural language, which compelled them to develop a personal understanding of the knowledge they were learning. For example, when solving the 'eight queens' puzzle, learners needed to understand the puzzle's rules and clearly explain their comprehension of the backtracking algorithm, including how to backtrack when encountering unplaceable queens on the chessboard. They also had to instruct the ChatGPT agent to follow their reasoning. This result aligns with previous research on learning by teaching, which showed when students explain concepts to their peers, they engage in deeper cognitive processes, thereby improving their own learning outcomes \cite{torshizi2019explain}. 
Our findings contribute to a deeper understanding of whether learners can effectively teach a teachable agent using natural language. Due to technological limitations, previous studies have rarely investigated teachable agents that allow learners to teach using natural language as if they were interacting with real peers \cite{lachner2022learning}. In most studies involving teachable agents, students are usually limited to specific tasks, such as completing graphical representations or answering multiple-choice questions, rather than engaging in natural discourse (e.g., \cite{chin2010preparing, biswas2016design, silvervarg2021teachable}). This is significant because, as previous research has highlighted, learning by teaching is fundamentally a social act, with its effectiveness being largely contingent on the degree of interactivity involved \cite{kobayashi2019learning}. Our findings demonstrated that when learners engage in teaching through natural language to a ChatGPT agent, they can also achieve significant knowledge gains. This underscores the promising potential of natural language interactions in enhancing the overall learning experience with teachable agents.

\subsection{The effect of ChatGPT teachable agent on programming skills}

%下一段写为什么programming 能力也能获得提高，这和过去研究有什么不同
Our results revealed that learners in the experimental group (EG) achieved higher scores in programming tests related to code clearness and readability, while there were no significant differences in code correctness compared to the control group (CG). In our learning task, code clearness refers to whether learners provided a logical flow in their coding, and readability pertains to the ease with which the code could be understood, including the use of natural language comments. These qualities are recognized as markers of high-quality code \cite{borstler2023developers}. Our findings suggest that by guiding a ChatGPT teachable agent in code generation, learners were able to refine their coding logic and develop a deeper understanding of what constitutes easily readable code from the perspective of the reader, rather than just the programmer. This indicates that the process of instructing the ChatGPT agent helped learners to clarify their thought processes in coding, ultimately contributing to their development of more structured and comprehensible code. Moreover, this learning by teaching approach likely encouraged learners to prioritize the readability and clearness of their code, emphasizing the importance of writing code that is understandable to others \cite{tan2024collaborative}.

However, our findings indicate that students using ChatGPT agents for teaching did not show significant differences in code correctness compared to the control group (CG), with the CG actually achieving slightly higher scores in code correctness. In other words, learners who interacted with ChatGPT agents were likely to produce code with a variety of bugs in the testing phase. One possible explanation is that ChatGPT agents can be ruled towards generating fully correct code (part of the ChatGPT pre-training that forced to generate correct codes) when responding to learners' guidance \cite{ma2024teach}, which deprives learners of the opportunity to practice determining code correctness themselves, such as finding bugs in the code. In contrast, the CG, which required students to write their own code, provided more opportunities to practice identifying bugs in their code. Data from the backend auto-judging platform also revealed that the experimental group (EG) had an average of 1.95 code submission attempts, whereas the CG had an average of 2.90 attempts, indicating that the experimental group passed the auto-judgement with fewer attempts. This finding was echoed previous studies that although the latest ChatGPT models have strong code generation accuracy \cite{markel2023gpteach}, there is a need for precise control of the teachable agents' cognitive behaviors (e.g., knowledge levels and question-asking) to facilitate the intended learning experience \cite{jin2024teach,lu2024generative} . Our results indicate that for ChatGPT-based teachable agents, those that occasionally generate incorrect responses may better promote the practice of error correction among learners. In other words, more advanced language models like GPT-4, with their extensive knowledge, may be less effective in developing learners' practical skills in error correction compared to more modest models like GPT-3.5, which can occasionally generate errors for learners to correct.

\subsection{The effect of ChatGPT teachable agent on SRL abilities} 
%finish draft
For RQ3, this study provides empirical evidence that learning with ChatGPT teachable agent can significantly affect students' SRL in the experimental group, as measured by the pre-post differences in self-efficacy and cognitive strategies. Overall, we found that students' self-efficacy improved after they teaching ChatGPT teachable agent. As developed by \cite{bandura1977analysis}, self-efficacy refers to subject-specific confidence in one’s ability to succeed in this subject. In our experiment, students' guidance to ChatGPT could mostly successfully generate codes by less attempts to pass the online judgement. In other words, the ChatGPT decreased students' difficulties in programming solution codes when solving the ``eight queens puzzle", which improved their confidence in the programming subjects. Moreover, as previous study indicated, the teachable agent allowing students to act the role of an expert, boosting their self-esteem and confidence compared to traditional student role \cite{pareto2009learning}. Other studies also indicated that observing the teachable agent's improvement based on their own guidance can act as a form of social modeling \cite{biswas2005learning}. Seeing the agent succeed can mirror the learners' own potential for improvement, thereby boosting their belief in their capabilities \cite{nye2013social}. Those factors can be the reasons of students' self-efficacy improvement when they teaching the ChatGPT agent. 

We also found that students' cognitive strategies improved more in EG compared to CG. It is worth noting that, unlike previous teachable agents where students' behaviors and methods of instruction were pre-defined (e.g., \cite{biswas2005learning, matsuda2010simstudent}), interactions with the ChatGPT agent were different. In the latter case, students needed to plan how to solve tasks themselves. As \cite{zimmerman2011self} stated , planning is an essential component of SRL strategies, as it helps students break down a task into manageable parts. The teaching process with the ChatGPT agent could be particularly beneficial as it guides the agent on how to solve a learning task by appropriately breaking it down and formulating a plan. Previous studies have shown that the ability to divide a learning task into parts and manage these tasks in collaboration with a ChatGPT system is a crucial skill for effective human-AI interaction \cite{fui2023generative, ahmad2023towards}. Therefore, our findings on the improvement of cognitive strategies suggest that when students teach a teachable ChatGPT agent, they become more engaged and spontaneous in employing detailed SRL strategies. These strategies mirror real-world learning scenarios more closely, where students not only aware their agents' mistakes but also guide them using suitable SRL techniques and adjusting strategies as needed.

\subsection{Implications}
% draft
Our study explored the effect of ChatGPT-based teachable agents on students' learning outcomes, resulting in several implications for educators and researchers: First, our study indicated that the natural language teaching process based on ChatGPT can enhance students' knowledge gains, validating that natural dialogue with a teachable agent can be an effective way for students to learn by teaching. Our findings recommend that educators create tasks requiring students to teach or explain concepts to ChatGPT agents using natural language to promote deeper cognitive processes. Second, it is evident that the pedagogical design of using AI as teachable agents should not focus solely on the accuracy of AI-generated content but also on the learning processes it stimulates. Incorporating scenarios where learners must identify and correct errors, such as purposeful mistakes by the AI, can be beneficial. Our findings suggest focusing more on lower-level ChatGPT, which may introduce more opportunities for students to correct errors, when designing ChatGPT-based teachable agents. Third, unlike traditional teachable agents with predefined behaviors, ChatGPT's flexibility allows learners to independently plan and analyse how to solve tasks, akin to real-world scenarios, thus requiring detailed SRL strategies. In other words, the ChatGPT-based teachable agent could be consider as one of method of SRL intervention that can facilitate students' self-efficacy and more cognitive strategies used. Future research should explore how SRL strategies can be integrated into the interactions between students and ChatGPT-based teachable agents. This could include prompting students to set goals, monitor their understanding, and reflect on their teaching process, thus making the learning experience more intentional and metacognitive.

\section{Conclusions}\label{sec5}

Given the growing interest in incorporating generative AI such as ChatGPT into education to enhance the learning experience, there is still room for investigation into how to empower students to take on a more proactive role in their learning. Learning by teaching is one of an important pedagogical strategy for students to actively focus on proactive learning, while it faced many limitation like lack of suitable peers. Previous research has shown that using teachable agents can effectively support this learning process, but these systems often require extensive development and typically lack the ability to facilitate natural language dialogue, limiting socialized experiences of learning by teaching. Our study explored the use of ChatGPT as a teachable agent to support students' learning by teaching in programming education. The findings indicate that engaging in natural language dialogue with ChatGPT enhances students' knowledge gains and improves  their programming skills, particularly in writing readable and logically sound code. However, teaching ChatGPT did not significantly help students develop error-correction skills, possibly because ChatGPT tends to generate correct code, reducing opportunities for students to identify and fix errors. Additionally, students' SRL abilities increased, suggesting that the teaching process with ChatGPT provides a better environment for implementing SRL strategies, such as independent planning and problem-solving. Overall, the study demonstrates the high potential of ChatGPT as a teachable agent in supporting students' learning by teaching processes, paving the way for future research in this area.

However, our study also meet several limitations. First, our study lacks of collecting and analysing students' conversation data on the teaching process of learning by teaching with ChatGPT agent. For example, we did not analysis the quality of teaching message of students or the quality of response message of ChatGPT in influencing students learning outcomes. The detailed analysis could increase our understanding of the ability of ChatGPT-based teachable agent in raising questions, and understanding of designing suitable scaffolding to support students' teaching process when they faced problems in teaching. Second, our study was conducted with a relatively small sample size of participants, which may bring bias into our results. The findings of our study require careful consideration. Third, the study primarily focused on short-term learning outcomes without considering the long-term retention of knowledge and skills. This limitation suggests that further research should examine the enduring impacts of using ChatGPT as a teachable agent over longer periods to determine whether the initial learning gains are sustained and translate into long-term educational benefits. Fourth, the study did not differentiate between various student demographics such as age, educational background, and gender, which might influence how students interact with and benefit from the ChatGPT-based teachable agent. Future studies should consider these variables to better understand the differential impacts of AI teachable agents across diverse learner groups.

\bibliographystyle{unsrt}  
\bibliography{BJET}

\end{document}